# A computationally efficient approach to modeling contact problems and fracture closure using superposition method[*]

HanYi Wang, ShiTing Yi and Mukul. M. Sharma

Petroleum & Geosystems Engineering Department, The University of Texas at Austin, 200 E. Dean Keeton St., Stop C0300 Austin, TX 78712-1585, United States

**Abstract**

The shape of fractures in an elastic medium under different stress distributions have been well studied in the literature, however, the fracture closure process during unloading has not been investigated thoroughly. The fracture surface is normally assumed to be perfectly smooth so that only the width and stress intensity at the fracture tip are critical in the analysis. In reality, the creation of a fracture in a rock seldom produces smooth surfaces and the resulting asperities on the fracture surfaces can impact fracture closure. Correctly modeling this fracture closure behavior has numerous applications in structural engineering and earth sciences. In this article, we present an approach to model fracture closure behavior in a 2D and 3D elastic media. The fracture surface displacements under arbitrary normal load are derived using superposition method. The contact stress, deformation, and fracture volume evolution can be estimated in a computationally efficient manner for various fracture surface properties. When compared with the traditional integral transform methods used to model fracture closure, the superposition method presented in this study produces comparable results with significant less computation time. In addition, with the aid of parallel computation, large scale fracture closure and contact problems can be successfully simulated using our proposed dynamic fracture closure model (DFCM) with very modest computation times.

## 1. Introduction

Fracture widths in a solid elastic medium are governed by the stress field around the crack tip and by parameters that describe the resistance of the material to the imposed stress. Thus, the analysis of stresses near the crack tip constitutes an essential part of fracture mechanics. For brittle materials exhibiting linear elastic behavior, methods of elasticity are used to obtain stresses and displacements in cracked bodies. England and Green (1963) studied 2D crack problems using complex variable techniques and the crack displacement was expressed as a double integral over the fracture length. Sneddon (1946) investigated the distribution of stresses produced in the interior of an elastic solid by the opening of an internal crack under the action of pressure applied to its surface. The solutions are derived from Westergaard's stress function (1939) and the critical crack propagation pressure under uniform load is obtained based on the rate of energy dissipation that proposed by Griffith (1920). Shah and Kobayashi (1971) presented solutions of stresses and the stress intensity factor for an elliptical crack embedded in an elastic solid and subjected to complex pressure distributions expressible in terms of a 2D polynomial function. The foundation of this work is based on the potential function proposed by Segedin (1967). Bell (1979) derived the solution for stresses and displacements due to arbitrarily distributed normal and tangential loads acting on a circular crack in an infinite body, in the form of series of Bessel function integrals. Understanding the effect of stress and load conditions on crack geometry is fundamental to solving crack propagation problems in solid materials. These stresses can be a result of pressured liquid injection, such as magma driven dykes (Rubin 1995), sub-glacial drainage of water (Tsai and Rice 2010), underground $CO_2$ or waste repositories (Abou-Sayed et al. 1989) or intentionally induced hydraulic fractures for stimulating hydrocarbon reservoirs (Economides and Nolte 2000) or for geothermal energy exploitation (Nemat-Nasser et al. 1983). Although hydraulic fracturing can be dated back to the 1930s (Grebe and Stoesser, 1935), it has been in the last twenty years that this practice has become commonplace. Combining the analytic solutions of stresses and displacements with a material balance on the injected fluid, leads to the development of extensively used pressurized fracture propagation models, such as the KGD model (Geertsma and De Klerk 1969; Khristianovich and Zheltov 1955), the PKN model (Nordgren 1972; Perkins and Kern 1961), and the radial model (Abe et al. 1976). Besides using analytic methods (e.g., complex potential function method and the integral transform method) to relate load, stress and displacement, numerical methods such as discrete element methods (McClure et al. 2016; Sesetty and Ghassemi 2015; Zhao 2014), cohesive zone methods (Bryant et al. 2015; Wang et al. 2016), and extended finite element methods (Dahi-Taleghani and Olson, 2011; Gordeliy and Peirce, 2013; Wang 2015; Wang 2016) can be used to solve stress and displacement fields that are coupled with material balance and

---

lubrication equations to model fluid driven fracture propagation. Even though these numerical methods enable us to model fracture propagation with complex interactions under fully coupled physics, just as analytic models, they assume smooth crack surfaces. This assumption is legitimate for modeling fracture initiation and propagation when internal load exerts sufficient pressure on the fracture surface to prop it open and maintain a finite fracture aperture. However, during the unloading process, the fracture gradually closes and the stress acting normal to the fracture plane results in fracture apertures approaching the scale of the surface roughness, and fracture surfaces can no longer be treated as perfectly smooth.

With the assumption of a pressurized fracture with perfectly smooth surfaces imbedded in an elastic medium, the opposite surfaces will come into contact all at once when the internal pressure acting on fracture surfaces drops to zero, and the fracture is then completely mechanically closed. In reality, the creation of cracks seldom produces perfectly smooth surfaces. It is well established that the fracture properties of brittle heterogeneous materials such as rock, concrete, ceramics, wood, and various composites strongly depend on the microstructure and on the damage process of the material. Both these parameters are also at the source of the roughness of crack surfaces (Morel et al. 2000). This is especially the case for large scale subsurface fractures, such as those generated by hydraulic fracturing. Van Dam et al. (2000) presented scaled laboratory experiments on hydraulic fracture closure behavior. Their work shows that the roughness of the fracture surfaces appears as a characteristic pattern of radial grooves and it is influenced by the externally applied stress. They also observed up to a15% residual aperture long after shut-in. Fredd et al. (2000) used sandstone cores from the East Texas Cotton Valley formation to show sheared fracture surface asperities that had an average height of about 0.09 inches. Warpinski et al. (1993) reported fracture surface asperities of about 0.04 and 0.16 inches for nearly homogeneous sandstones and sandstones with coal and clay rich bedding planes, respectively. Field measurements (Warpinski et al. 2002) using a down-hole tiltmeter array indicated that the fracture closure process is a smooth, continuous one which often leaves 20%-30% residual fracture width.

Fractures play a decisive role in determining production and its decline trend. In natural fractures, the most important properties that affect the fluid flow are the fracture width, normal stress, contact area and the roughness of the fracture surfaces. These properties are all interdependent and directly affect each other. In propped fractures, the stress distribution after closure can substantially influence the crushing or embedment of proppants and the resulting conductivity (Warpinski 2010). In addition, the fracture closure behavior significantly impact our estimation and correct interpretation of in-situ stress through diagnostic fracture injection tests (McClure et al. 2016; Wang and Sharma 2017b) or pump-in and flow back tests (Raaen et al. 2001; Savitski and Dudley 2011). The changes of fracture compliance during closure is the key input parameter for flow back analysis (Fu et al. 2017).

An extensive literature exists on the relationship between fracture aperture, conductivity, internal pressure, and effective stress. These studies include analytical models (Greenwood and Williamson, 1966; Gangi, 1978; Brown and Scholz, 1985; Cook, 1992; Adams and Nosonovsky, 2000; Myer, 1999) and numerical approaches (Pyrak-Nolte and Morris, 2000; Lanaro, 2000; Kamali and Pournik 2016) as well as experimental studies (Barton et al. 1985; Brown et al. 1998; Marache et al. 2008; Matsuki et al. 2008). However, all these studies focus on detailed investigations of the closure of fractures modeled as two parallel plates with rough surfaces, and ignore the impact of the fracture geometry and the evolution of the stress re-distribution during fracture closure.

Even though analytic methods can give closed from solutions for fracture displacement and stress distribution under arbitrary load, it requires a priori knowledge of load distribution (i.e., the arbitrary load can be written as a function of local coordinates). However, in contact problems, the total pressure/stress acting on the opposite surfaces of fracture is a dynamic process and coupled with fracture displacement, so it is not known in advance and does not have a general form that can be represented by smooth analytic functions. To resolve this issue, Wang and Sharma (2017a) presented an integral transform method to model contact problem with analytic solutions that relate load and displacement, the solution integrals have to be divided into piecewise segments to approximate the distribution of possible discrete internal loads, however, this method is not very efficient because the number of integrals that need to be evaluated grows exponentially with increasing number of segments. Using numerical methods (such as finite element method, finite volume method, boundary element method, etc.) to model fracture closure and contact problems also face challenges, because in many cases there is a sharp contrast of scales between fracture geometry (e.g., tens or hundreds of meters for subsurface fractures) and surface asperities (normally in the range of μm or mm). For a numerical model to capture both large scale fracture deformation and small scale contact behavior during fracture closure, the mesh or element size has to be extremely small (equivalent to the scale of the fracture aperture), which makes this effort very computationally expensive. In addition, contact problems are often non-linear and require a large number of iterations to converge, this makes numerical modeling of fracture closure processes more challenging, especially when the fracture size is large or a large number of fractures need to be modeled simultaneously.

In this article, we propose a new method and general algorithms to model the dynamic behavior of fracture closure on rough fracture surfaces and asperities. Analytic solutions that relate fracture width with arbitrary normal net pressure/stress using superposition are derived. Coupling analytic solutions with a general contact law that up-scales surfaces roughness and asperities, large scale 2D and 3D fracture closure behavior can be modeled efficiently. Different methods of modeling fracture closure behavior and their computation efficiency are also compared and investigated.

## 2. Fracture Surface Displacement with Arbitrary Normal Load

When the internal pressure of a fracture declines, the fracture aperture diminishes accordingly until the fracture faces come into contact when the asperities of opposing fracture walls touch each other. This induces an additional normal load (i.e., contact stress) acting on the area of contact and changes the subsequent fracture closure behavior as the internal pressure declines further. Here we are not seeking a general solution that relates displacement to the load, because the load itself is dynamically related to fracture geometry, material properties and surface roughness during the fracture closure process. Instead, we determine the displacement based on superposition, by adding the influence of discretized fracture segment/surface areas that subject to constant computed loads. In this section, we present a method to determine fracture aperture under an arbitrary normal load in a two and three dimensional elastic medium using superposition.

### 2. 1 Two Dimensional Fracture

The problem of a pressurized line crack in an infinite, elastic domain is defined by the following conditions:

$$\sigma_{xy} = 0, \quad |x| < a, y = 0 \tag{1}$$
$$\sigma_{yy} = -p(x), \quad |x| < a, y = 0 \tag{2}$$
$$u_y = 0, \quad |x| \geq a, y = 0 \tag{3}$$

The fracture in Cartesian coordinate x-y is shown in **Fig.1**. $p(x)$ is the pressure acting on the crack surface. All displacements, u, and stresses, σ, are zero at infinity. Linear elasticity and plane strain are assumed.

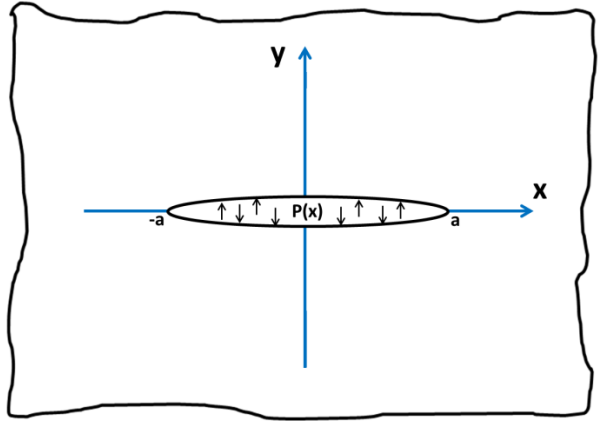

**Fig.1 2D Crack ($|x| < a$) in an infinite domain**

The stress equilibrium equations in a 2D domain can be expressed as

$$\begin{cases} \dfrac{\partial \sigma_x}{\partial x} + \dfrac{\partial \sigma_{xy}}{\partial y} = 0 \\ \dfrac{\partial \sigma_y}{\partial y} + \dfrac{\partial \sigma_{xy}}{\partial x} = 0 \end{cases} \tag{4}$$

These equations are satisfied if we assume

$$\begin{cases} \sigma_x = \dfrac{\partial^2 \Phi}{\partial y^2} \\ \sigma_y = \dfrac{\partial^2 \Phi}{\partial x^2} \\ \sigma_{xy} = -\dfrac{\partial^2 \Phi}{\partial x \partial y} \end{cases} \tag{5}$$

where $\Phi$ is termed the Airy Stress Function. Hooke's Law that relate stress and strain:

$$\begin{cases} E\varepsilon_x = \sigma_x - \upsilon(\sigma_y + \sigma_z) \\ E\varepsilon_y = \sigma_y - \upsilon(\sigma_x + \sigma_z) \\ E\gamma_{xy} = 2(1+\upsilon)\sigma_{xy} \end{cases} \tag{6}$$

where $E$ is Young's modulus and $v$ is Poisson's ratio, and $\varepsilon, \gamma$ are normal and shear strain respectively. For plane stress condition, $\sigma_z = 0$ and for plane strain condition, $\sigma_z = v(\sigma_x + \sigma_y)$.

A convenient equation that represent three strains are defined in terms of derivatives of only two displacements is given by

$$\frac{\partial^2 \varepsilon_x}{\partial y^2} + \frac{\partial^2 \varepsilon_y}{\partial x^2} = \frac{\partial^2 \gamma_{xy}}{\partial x \partial y} \tag{7}$$

Substituting Eq.(5) into Eq.(6), and using Eq.(7), the stress equilibrium can be expressed as,

$$(\nabla^4 \Phi) = 0 \tag{8}$$

Westergaard (1939) found an Airy Stress Function of complex numbers can be the solution for the stress field in an infinite plate containing a crack. He discussed several Mode I crack problems that could be solved using:

$$\Phi = Re\{\bar{\bar{Z}}(z)\} + yIm\{\bar{Z}(z)\} \tag{9}$$

where $Re$ and $Im$ denote the real and imaginary part, and

$$\bar{Z}(z) = \frac{d(\bar{\bar{Z}}(z))}{dz} \tag{10}$$

$$Z(z) = \frac{d(\bar{Z}(z))}{dz} \tag{11}$$

$z$ is a complex number. Then, the stresses can be calculated by taking derivatives of the Airy Stress Function:

$$\begin{cases} \sigma_x = ReZ - yImZ' \\ \sigma_y = ReZ + yImZ' \\ \sigma_{xy} = -yReZ' \end{cases} \tag{12}$$

Where,

$$Z'(z) = \frac{d(Z(z))}{dz} \tag{13}$$

For plane strain and symmetric loading on y=0, $\sigma_{xy} = 0$, and the displacement field at x-direction, $u_x$, and y-direction, $u_y$, can be determined:

$$\begin{cases} u_x = \frac{(1 - 2v)(1 + v)Re(\bar{Z}) - y(1 + v)Im(Z)}{E} \\ u_y = \frac{2(1 - v^2)Im(\bar{Z}) - y(1 + v)Re(Z)}{E} \end{cases} \tag{14}$$

The use of Airy Stress Function is a powerful technique for solving 2D equilibrium elasticity problems. After finding the Airy Stress Function for a specific problem, all the stress field and displacement can be determined. Considering a case of normal traction p from x=b to c is applied to a line crack $|x| \leq a$ in an infinite elastic body (shown in **Fig.2**). The complex number Z for Westergaard solution for Mode I fracture is given by (Tara et al. 1973):

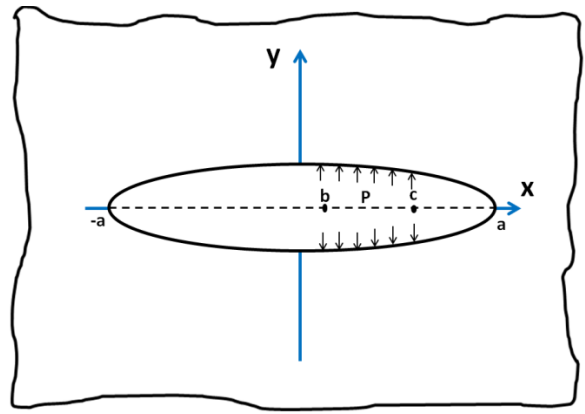

**Fig.2 Partial load on a 2D line crack with a constant pressure p applied on $b \leq x \leq c$**

$$Z(z) = \frac{1}{\pi}p \left\{ \begin{array}{c} \sin^{-1}\dfrac{a^2 - cz}{a(z-b)} - \sin^{-1}\dfrac{a^2 - bz}{a(z-b)} \\ + \dfrac{\sin^{-1}\dfrac{c}{a} - \sin^{-1}\dfrac{b}{a}}{\sqrt{1 - \left(\dfrac{a}{z}\right)^2}} - \dfrac{\sqrt{a^2 - c^2} - \sqrt{a^2 - b^2}}{\sqrt{z^2 - a^2}} \end{array} \right\} \quad (15)$$

And the imaginary part of the first derivative of $Z$ over z is:

$$\text{Im}(\bar{Z}) = \frac{1}{\pi}p \left\{ \begin{array}{c} (c-x)\cosh^{-1}\dfrac{a^2-cx}{a|x-c|} - (b-x)\cosh^{-1}\dfrac{a^2-bx}{a|x-b|} \\ +(\sin^{-1}\dfrac{c}{a} - \sin^{-1}\dfrac{b}{a})\sqrt{a^2-x^2} \end{array} \right\} \quad (16)$$

Finally, the width of a fracture, $w_f$, along x-direction can be calculated by set y=0 in Eq.(14):

$$w_f(x|y=0) = 2u_y(x|y=0) = \frac{4(1-v^2)}{E}\text{Im}(\bar{Z}) = \frac{4}{E'}\text{Im}(\bar{Z}) \quad (17)$$

where $E'$ is plane strain Young's modulus.

For a special case, when b=-a and c=a, Eq.(17) is reduced to the analytical solution for a constant internal pressure:

$$w_f(x) = \frac{4p}{E'}\sqrt{a^2 - x^2} \quad (18)$$

If the fracture from –a to a is divided into a total number of n line segments, and in the i$^{th}$ ($1 \leq i \leq n$) segment, there exists a uniform pressure $p_i$ that acts on the opposite fracture surface. If n is large enough, the pressure distribution can be approximated to any distribution of p(x). The solution of fracture width for each individual segment with constant pressure is given in Eq.(17). Use superposition by adding the influence of each individual segment over the entire fracture, the final fracture width profile for any given pressure distribution can be obtained:

$$w_f(x) = \sum_{i=1}^{n} w_f\left(x | x \in i^{th} \text{ segment}, p|p = p_i\right) \quad (19)$$

## 2. 2 Three Dimensional Fracture

For the three-dimensional case assume that a crack is created in the interior of an infinite elastic medium and that it is "penny-shaped" with a radius of R, as shown in **Fig.3**. The fracture is axially symmetry about the z-axis, and the internal pressure is a function a radius r. In the absence of body force, the equations of elastic equilibrium in a cylindrical coordinate system reduce to (Love 1934):

$$\frac{\partial \sigma_r}{\partial r} + \frac{\partial \sigma_{rz}}{\partial z} + \frac{\sigma_r - \sigma_\theta}{r} = 0 \quad (20)$$

$$\frac{\partial \sigma_{rz}}{\partial r} + \frac{\partial \sigma_z}{\partial z} + \frac{\sigma_{rz}}{r} = 0 \quad (21)$$

At the fracture surface ($r \leq R$), the boundary conditions are:

$$\sigma_{rz} = 0 \quad (22)$$

$$\sigma_z = -p(r) \quad (23)$$

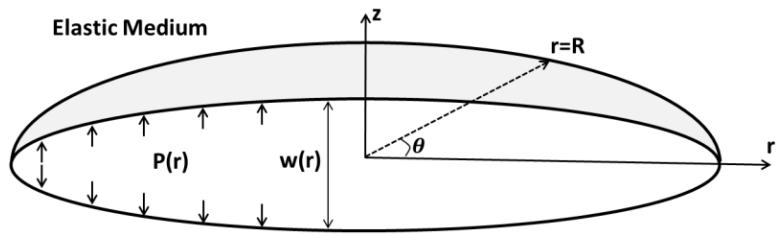

**Fig.3 Penny shaped fracture in cylindrical coordinates.**

Through integral transforms, the fracture aperture for a given internal pressure, p(r), can be obtained as a double integral (Sneddon 1946):

$$w_f(r) = \frac{8}{\pi E'} \int_r^R \frac{1}{\sqrt{\xi^2 - r^2}} \int_0^\xi \frac{p(r)}{\sqrt{\xi^2 - r^2}} r \, dr \, d\xi \quad (24)$$

where $\xi$ is a dummy variable, $w_f(r)$ is fracture width along r-direction. Eq.(24) is not suitable for contact problems because p(r) is itself coupled with w(r), which is not known in the first place. To determine $w_f(r)$ based on superposition, we first need the solution of w(r) when p(r) is constant over a circular area. Let's assume p(r) is imposed within a radius a:

$$p(r) = \begin{cases} p & (0 < r \le a) \\ 0 & (a < r \le R) \end{cases} \quad (25)$$

Let $G(\xi)$ be the inner integral of Eq.(24):

$$G(\xi) = \int_0^\xi \frac{p(r)}{\sqrt{\xi^2 - r^2}} r \, dr \quad (26)$$

Eq.(26) has to be evaluated piecewise for different ranges of r. When $0 < r \le a$

$$G(\xi)_{0<r\le a} = \int_0^\xi \frac{p}{\sqrt{\xi^2 - r^2}} r \, dr = p\xi \quad (27)$$

When $a < r \le R$

$$G(\xi)_{a<r\le R} = \int_0^a \frac{p}{\sqrt{\xi^2 - r^2}} r \, dr + \int_a^u \frac{p}{\sqrt{\xi^2 - r^2}} r \, dr = p(\xi - \sqrt{\xi^2 - a^2}) \quad (28)$$

The outer integral of Eq.(24) also needs to be evaluated in a piecewise manner. When $0 < r \le a$

$$w_f(r)_{0<r\le a} = \frac{8}{\pi E'} \int_r^a \frac{G(\xi)_{0<r\le a}}{\sqrt{\xi^2 - r^2}} d\xi + \frac{8}{\pi E'} \int_a^R \frac{G(\xi)_{a<r\le R}}{\sqrt{\xi^2 - r^2}} d\xi = \frac{8p}{\pi E'} (\sqrt{R^2 - r^2} - \int_a^R \frac{\sqrt{\xi^2 - a^2}}{\sqrt{\xi^2 - r^2}} d\xi) \quad (29)$$

When $a < r \le R$

$$w_f(r)_{a<r\le R} = \frac{8p}{\pi E'} \int_r^R \frac{G(\xi)_{a<r\le R}}{\sqrt{\xi^2 - r^2}} d\xi = \frac{8p}{\pi E'} (\sqrt{R^2 - r^2} - \int_r^R \frac{\sqrt{\xi^2 - a^2}}{\sqrt{\xi^2 - r^2}} d\xi) \quad (30)$$

Eq.(29) and (30) give the fracture width distribution under the load given by Eq.(25). The integrals in the solutions can't be evaluated analytically without resorting to elliptic integrals of the second kind (Abramowitz and Stegun 1972). With the above derivation, the solution of fracture width as a function of fracture radius, for a constant pressure applied in the circular region of $a \le r \le b$, can be obtained using superposition, as shown in **Fig.4**.

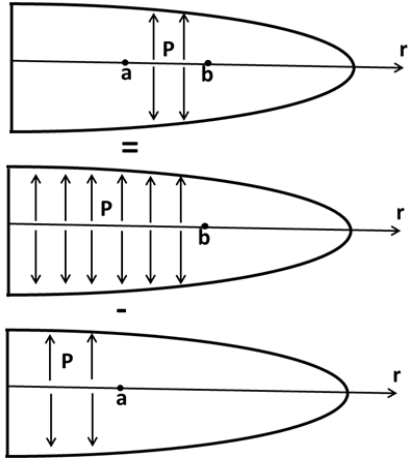

Fig.4 Solution of a constant pressure applied on $a \leq r \leq b$ using superposition.

Using the solution of the constant pressure acting on $0 \leq r \leq b$ minus the solution of the constant pressure acting on $0 \leq r \leq a$, we get the fracture width distribution for a constant pressure acting on $a \leq r \leq b$:

$$w_f(r) = \begin{cases} \dfrac{8p}{\pi E'}\left(\displaystyle\int_a^R \dfrac{\sqrt{\xi^2-a^2}}{\sqrt{\xi^2-r^2}}d\xi - \displaystyle\int_b^R \dfrac{\sqrt{\xi^2-b^2}}{\sqrt{\xi^2-r^2}}d\xi\right) & \text{for } 0 \leq r < a \\ \dfrac{8p}{\pi E'}\left(\displaystyle\int_r^R \dfrac{\sqrt{\xi^2-a^2}}{\sqrt{\xi^2-r^2}}d\xi - \displaystyle\int_b^R \dfrac{\sqrt{\xi^2-b^2}}{\sqrt{\xi^2-r^2}}d\xi\right) & \text{for } a \leq r < b \\ \dfrac{8p}{\pi E'}\left(\displaystyle\int_r^R \dfrac{\sqrt{\xi^2-a^2}}{\sqrt{\xi^2-r^2}}d\xi - \displaystyle\int_r^R \dfrac{\sqrt{\xi^2-b^2}}{\sqrt{\xi^2-r^2}}d\xi\right) & \text{for } b \leq r < R \end{cases} \quad (31)$$

For a special case, when a=0 and b=R, Eq.(31) reduces to the well-known analytical solution of a penny-shaped crack for a constant internal pressure:

$$w_f(r) = \frac{8p}{\pi E'}\sqrt{R^2-r^2} \quad (32)$$

**Fig.5** shows a penny-shaped fracture discretized into a total number of n circular segments along the fracture radius. In each segment, there exists a uniform pressure that acts on the opposite fracture surface. If n is large enough, the pressure distribution can be arbitrary, $p(r)$.

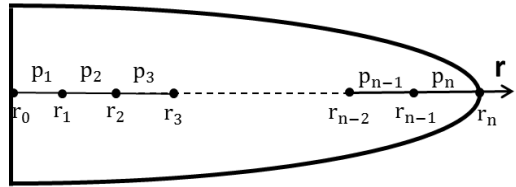

Fig.5 Illustration of discretizing fracture radius into a number of segments with uniform pressure within each segment.

The solution for fracture width for an individual segment is given in Eq.(31). Again, we use the superposition principle by adding the influence of each individual segment over the entire fracture to get the final solution of fracture width for any given pressure distribution:

$$w_f(r) = \sum_{i=1}^n w_f\left(r | r_{i-1} \leq r < r_i, p | p = p_i\right) \quad (33)$$

There are two main advantages of using superposition. One is that the mathematical form of the pressure distribution, $p(r)$, is not required *a priori*, the other is the pressure distribution does not necessarily need to be continuous, which suits the nature of contact problems.

## 3. Solution Dependent Contact Load

If the fracture surface is perfectly smooth, according to Eq.(18) and Eq.(32), the entire fracture surface area will come into contact all at once when the internal pressure drops to zero. However, as discussed previously, depending on material properties, heterogeneity and loading conditions, the generated crack or fracture surface is always rough with asperities distributed across the surface area. So fractures can retain a finite aperture after mechanical closure due to a mismatch of asperities on the fracture walls, this is especially the case for subsurface fractures that are stimulated in hydrocarbon and geothermal reservoirs. Scaled laboratory experiments show that the created fracture surfaces are rough (shown in **Fig. 6**)

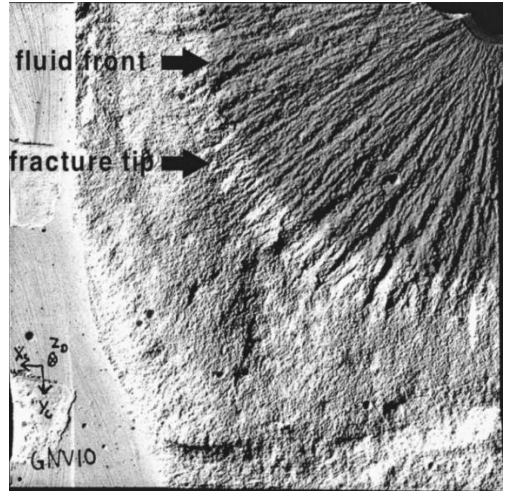

**Fig.6 Example of fracture surface roughness pattern generated by a hydraulic fracture (van Dam et al. 2000)**

Detailed measurements and modeling of surface roughness and mechanical properties of asperities for every fracture encountered is not practical and, many times not even possible, so it is desirable to study and investigate specimen samples, and upscale the influence of surface microscopic structure to some macroscopic empirical relationships (i.e., a contact law) that relates fracture width and the associated contact stress. Zangerl et al. (2008) compiled laboratory and in-situ experiments of fracture closure behavior for different rock samples and showed that the effective normal stress and fracture normal closure (the displacement from zero effective normal stress condition) is highly non-linear, as shown in **Fig.7**.

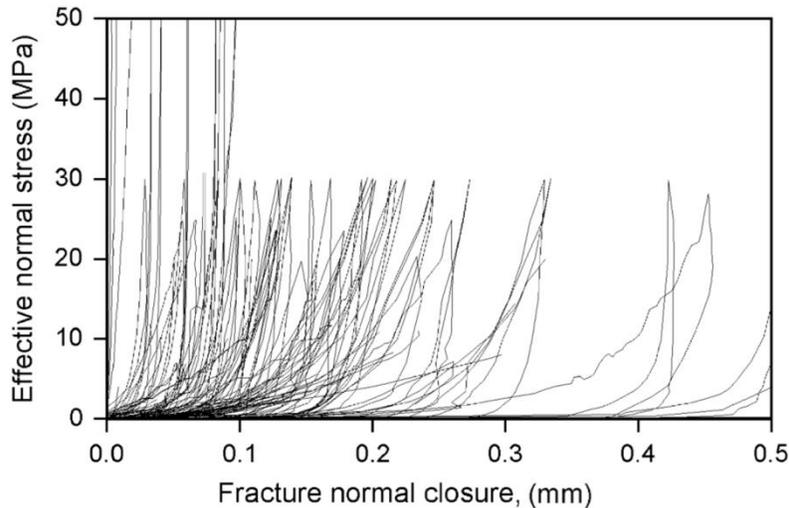

**Fig.7 Compiled laboratory and in-situ experiments showing highly non-linear normal closure behavior (Modified from Zangerl et al. 2008)**

To quantify the non-linear fracture closure behavior, Willis-Richards et al. (1996) proposed a relationship to estimate contact stress for a certain fracture aperture, based on the work of Barton et al. (1985):

$$\sigma_c = \frac{\sigma_{ref}}{9}\left(\frac{w_0}{w_f} - 1\right) \text{ for } w_f \leq w_0 \tag{34}$$

$\sigma_c$ is the contact stress acting normal on the fracture, and $\sigma_{ref}$ is the effective normal stress at which the aperture is reduced by 90%. $w_f$ is the fracture aperture and, $w_0$ is the fracture aperture when the contact stress is equal to zero. When $w_f > w_0$, the contact stress is zero because the fracture walls are mechanically detached. It should be emphasized that the contact width $w_0$ is determined by the tallest asperities, and the strength, spatial and height distribution of asperities are reflected by the contact reference stress $\sigma_{ref}$ (e.g., if the tallest asperities on two fracture samples are the same, then they should have the same $w_0$, but the one with a higher median asperity height or higher Young's modulus will have a higher value of $\sigma_{ref}$, provided other properties are the same). The influence of all the rock properties, modulus, asperities patterns, density, distribution, etc., can be up-scaled into two contact parameters, $w_0$ and $\sigma_{ref}$, using Eq.(34) to match the (e.g., curves in Fig.7) displacement vs loading stress curves for fractured rock samples

## 4. Coupling Superposition Solutions with Contact Law to Model Fracture Closure

In this section, we present the algorithm to determine fracture aperture and contact stress distribution with a given fracture geometry, rock properties, fracture internal pressure and parameters for the contact law. When the fracture closes on rough surfaces and asperities as the internal pressure drops, the total pressure acting on the fracture surfaces becomes the internal pressure plus contact stress. The total pressure distribution impacts surface displacement and in turn, determines the contact stress, so the contact stress and fracture width distribution need to be solved simultaneously. **Fig.8** shows a schematic of the algorithm used for solving the fracture contact problem based on a Picard iteration (Berinde 2004). During the first step, if the contact stress is not known, it is assumed to be zero, and the resulting fracture width is smaller than the actual width if some portion of the fracture tip is already in contact. Based on this initial guess of the fracture width, we can calculate the corresponding contact stress and get the updated total pressure distribution. Convergence is achieved when the absolute change of total pressure between two consecutive iterations is smaller than a specified tolerance. The total net pressure/stress, $p_{total}$, acting on the fracture surface is determined by:

$$p_{total} = p_f + \sigma_c - \sigma_{min} \qquad (35)$$

where $p_f$ is fluid pressure inside the fracture, $\sigma_{min}$ is the far field stress that is perpendicular to the fracture face. From Eq.(34), we can observe that the relationship between contact stress and fracture aperture is highly non-linear, and small changes of fracture aperture can lead to very large increases in contact stress. This poses great challenges for a coupled iterative model. The Picard relaxation factor, $\alpha_r$, avoids over correction in each iteration, and has an value between 0 and 1. The proposed algorithm of modeling fracture closure behavior is a general algorithm, where different contact laws and different methods that relate displacement and loading conditions can be applied.

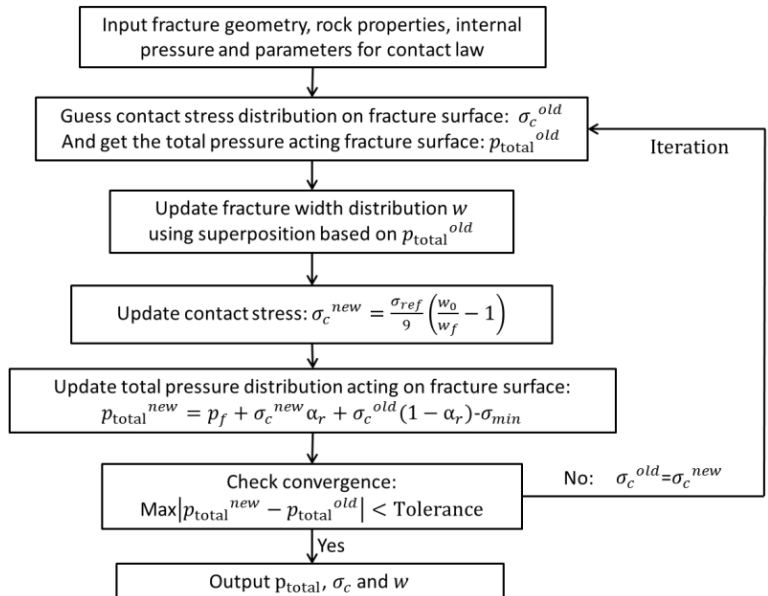

Fig.8 Solution flow diagram for solving coupled contact stress and fracture width distribution.

## 5. Validation of Superposition Implementation

In this section, we examine and validate the superposition method implemented in this study by comparing the fracture width distribution with analytic solutions under uniform internal pressure. The fracture is divided into a number of segments along the fracture length or fracture radius, and a pressure is imposed on each segment to represent a uniform internal pressure condition. First, a 2D plan strain case is examined as shown in **Fig.9**. The fracture length is 50 m, Young's modulus is 20 GPa, Poisson's ratio is 0.25 and the internal pressure is 5 MPa. The results from the superposition method, using Eq.(19), with a different number of segments are compared with the analytic solution, using Eq.(18). It can be observed that when the total number of segment is only 6 or 8, the superposition method overestimates fracture width, but when the total number of segment increases to 10, the results from the superposition method agree very well with the analytical solution. Next, we examine a 3-D penny-shaped fracture with the same Young's modulus, Poisson's ratio and internal pressure as the 2-D case, and set the fracture radius to be 50 m. The results from superposition method, using Eq.(33), with different numbers of segments are compared with the analytical solution, using Eq.(32), are shown in **Fig.10**. Results from the superposition method match very well with the analytical solution. It seems that the superposition method for a 3-D penny-shaped fracture is less sensitive to the total number of segments or segment size than the 2-D fracture.

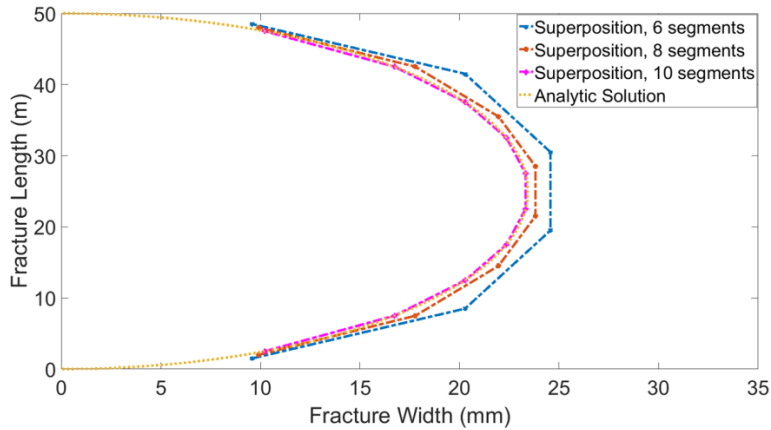

Fig.9 Validation of superposition implementation for 2D fracture with different numbers of segments.

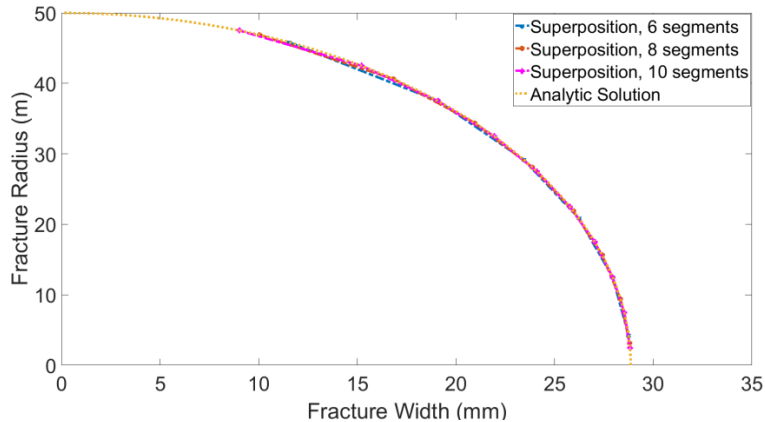

Fig.10 Validation of superposition implementation for 3-D fracture with different numbers of segments.

## 6. Fracture Closure and Contact Stress

In this section, the dynamic fracture closure process with declining internal pressure and the impact of surface roughness properties (represented by up-scaled contact parameters $w_0$ and $\sigma_{ref}$) will be examined for both 2-D and 3-D fracture geometries.

### 6. 1  2-D Fracture Closure

#### 6. 1.1 Uniform Far-Field Stress

A Base Case was set up to demonstrate the method. The fracture has a tip-to-tip length of 20 m (a =10 m in Fig.2) with a far-field stress $\sigma_{min}$ of 20 MPa. The surrounding rock has a Young's modulus of 20 GPa, Poisson's ratio of 0.25. On the fracture surface, the contact width $w_0$ is 2 mm and the contact reference stress $\sigma_{ref}$ is 5 MPa. The initial fluid pressure inside the fracture is 25 MPa. The length of the discretized segment for all the following simulated cases is set as 0.2 m. Using the proposed algorithm (Fig.8), the evolution of the fracture width profile and contact stress distribution can be determined, as the fluid pressure inside the fracture gradually declines. The results are shown in **Fig.11** and **Fig.12**. To demonstrate the impact of fracture roughness and surface asperities on fracture closure behavior, the case without surface asperities (fracture surface is completely planar and smooth) is also included. The result shows that at a relatively high fracturing fluid pressure, the fracture asperities have negligible impact on fracture width distribution, and the contact stress is always concentrated at the tip of the fracture, where the contact stress is much higher than in the middle of the fracture. We can also see that the fracture surfaces do not contact each other like parallel plates. In fact, the fracture closes on rough surfaces starting from the tip, and closes progressively all the way from the edges to the center of the fracture. As the fluid pressure continues to decline, more and more of the fracture surfaces come into contact and this changes the subsequent fracture closure behavior. At lower fluid pressures, contact stresses start to counter-balance the far-field stress and the fracture becomes stiffer and less compliant. If the fracture faces were perfectly parallel and smooth, the fracture width would have collapsed to zero when the fluid pressure dropped to 20 MPa (i.e., the far-field stress). However, because of the surface asperities and mismatches, the fracture still retains a residual fracture width even when the fluid pressure inside the fracture drops below the far-field stress.

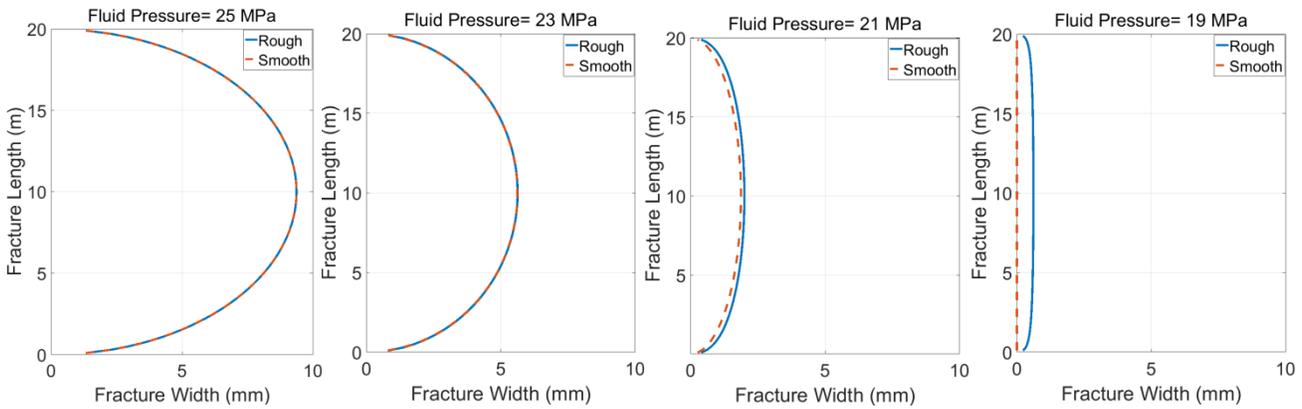

**Fig.11** Fracture width evolution with and without surface asperities at different fluid pressure under uniform far-field stress.

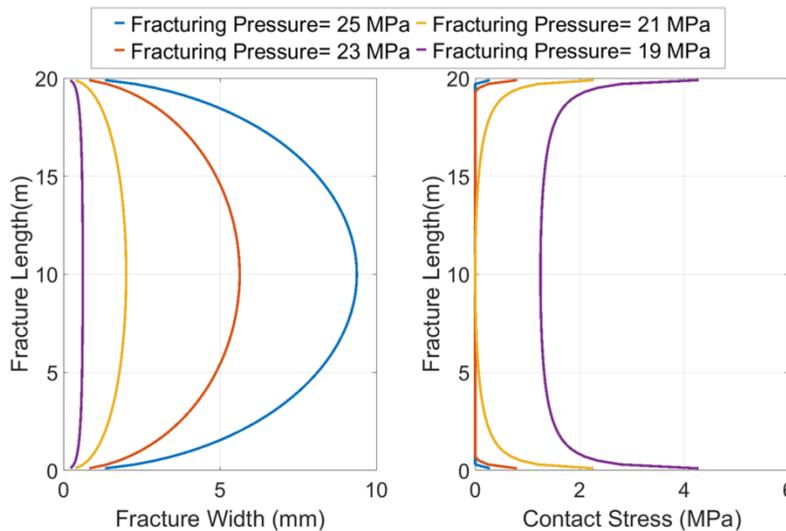

**Fig.12** Fracture width and the corresponding contact stress distribution at different fluid pressure under uniform far-field stress.

**Fig.13** shows the fracture width and the corresponding contact stress for different values of contact width $w_0$ with a contact reference stress $\sigma_{ref}$ of 5 MPa using Eq.(34). As expected, when the fracture width is larger than or equal to the contact width, the contact stress is zero. However, when the fracture width near the fracture tip is smaller than the contact width, the contact stress and fracture width follow a non-linear relationship, resembles that of Fig.7. As the fracture width gets smaller and smaller, more force is needed to further close the fracture (less compliant fracture). **Fig.14** shows the evolution of fracture average width for different contact width while other input parameters remain the same as the Base Case. As can be observed, when the pressure is relatively high (above 23 MPa), the fracture average width and fluid pressure follows a linear relationship, since the impact of asperities is negligible. However, as the pressure continues to decline, a larger contact width leads to an earlier departure from the linear relationship. This is because the rough fracture walls will come into contact at a higher pressure if the contact width is larger, and the larger contact width results in larger residual fracture width when the fluid pressure drops to 15 MPa.

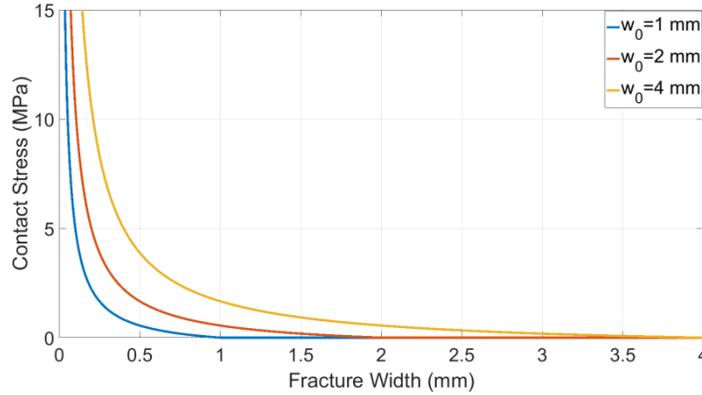

**Fig.13 The relationship between contact stress and fracture width for different $w_0$.**

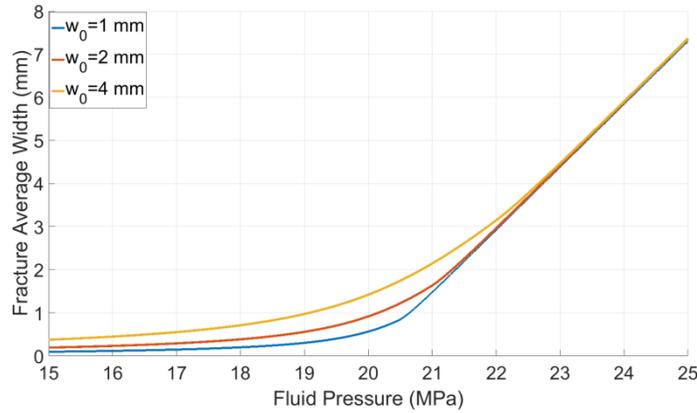

**Fig.14 Fracture average width evolution for different $w_0$ during closure with 2D fracture geometry**

**Fig.15** shows the fracture width and the corresponding contact stress for different values of $\sigma_{ref}$ with $w_0$ of 2mm using Eq.(34). It can be observed that for the same contact width, the higher the contact reference stress, the more rapid the increase of contact stress as the fracture width decreases. Physically, the contact reference stress represents how hard and compliant the fracture surface asperities are. **Fig.16** shows the evolution of fracture average width for different contact reference stress while other input parameters remain the same as the Base Case. The results reveal that the contact reference stress does not have much impact on the pressure at which the fracture asperity and pressure depart from a linear relationship, but it does impact the fracture width evolution after most of fracture surface has come into contact. The lower the contact reference stress, the smoother the change in fracture average width as pressure declines. A lower contact reference stress also leads to smaller residual fracture width.

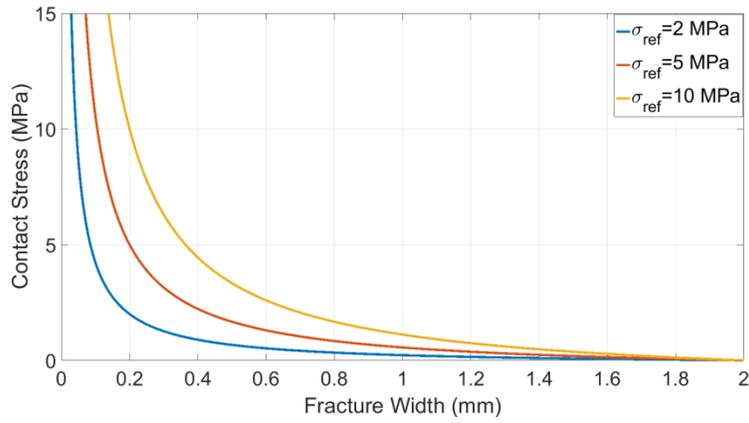

Fig.15 The relationship between contact stress and fracture width for different $\sigma_{ref}$.

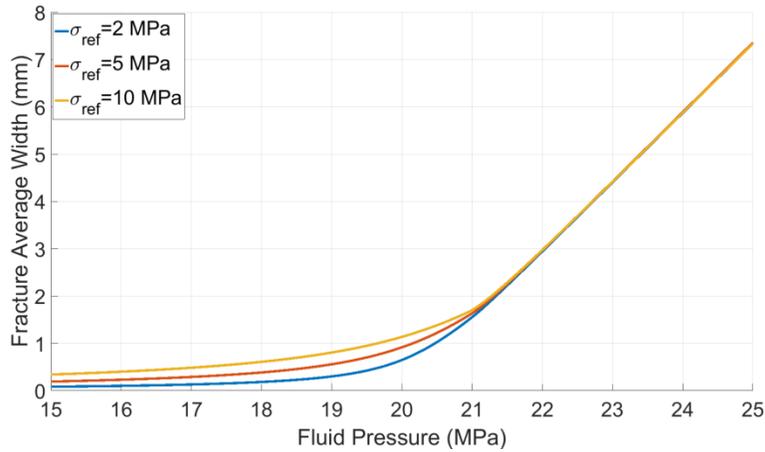

Fig.16 Fracture average width evolution for different $\sigma_{ref}$ during closure with 2D fracture geometry.

There is compelling field evidence to indicate that hydraulic fracture closure occurs on rough fracture surfaces in the field. **Fig.17** shows the normalized tilt-meter response plotted against wellbore pressure during the shut-in period of well-2B at three different stations from the GRI/DOE M-site. Each tilt-meter response is normalized by dividing the maximum tilt (displacement, also an indication of fracture width) measured at that instrument during the test. The tilt-meter demonstrates that soon after shut-in, the measured displacement declines linearly with pressure (roughly constant fracture stiffness). After the wellbore pressure declines to a certain level, the measured displacement vs pressure departs from a linear relationship. Even though these data were measured after weeks of shut-in, we can still observe the existence of a residual fracture width supported by surface asperities and mismatches even after the fracturing fluid pressure drops below the minimum in-situ stress. This field measurement is consistent with the general trend shown in Fig.14 and Fig.16. If the fracture geometry and rock properties can be obtained with certainty, the tilt-meter measurement, along with diagnostic fracture injection tests (DFITs), can be used to infer the contact parameters (i.e., $w_0$ and $\sigma_{ref}$) using our proposed model. The 2D fracture closure model presented here can be applied to both PKN and KGD fracture geometry, the difference is that for a PKN type fracture geometry (fracture length ≥ fracture height), a plane strain condition is assumed in the direction of fracture length, while for KGD type fracture (fracture height ≥ fracture length), the plane strain condition is assumed in the direction of fracture height.

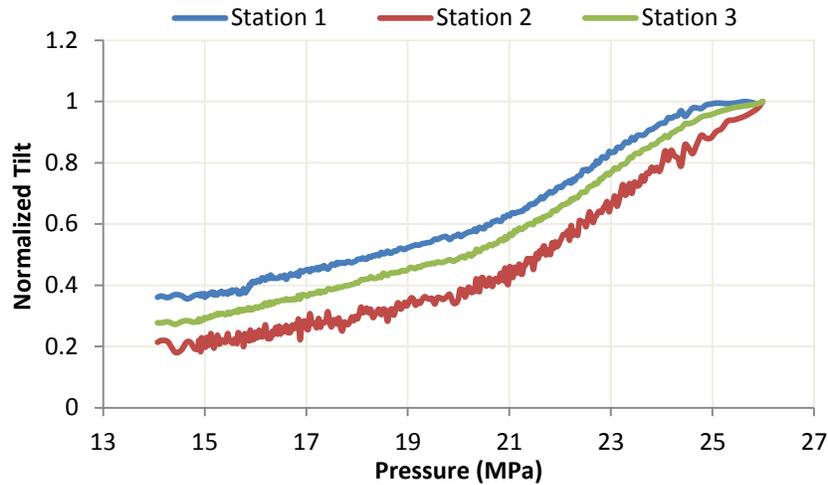

Fig.17 Normalized tilt-meter data from shut-in of Well-2B from the GRI/DOE M-site. Data Courtesy of Norm Warpinski.

### 6. 1.2 Non-Uniform Far-Field Stress

2D fracture models have been used for decades with reasonable success. However, in certain formations, neither of these two models (KGD and PKN) can be used successfully to model hydraulic fracture propagation. Today, pseudo-three-dimensional (P3D) models are the most commonly used hydraulic fracture model. P3D models, are in essence an extension of the classical PKN, but unlike the PKN model, the height of the P3D fracture is not limited to the reservoir thickness, as shown in **Fig.18**, the fracture is allowed to grow vertically into the adjacent layers, and the fracture width distribution is still based on the 2D plane strain conditions (no strain in the x-direction) of the PKN model. Using the superposition method of Eq.(19), the fracture width distribution under an arbitrary far-field stress can be obtained by first using Eq.(35) to get the total net pressure/stress distribution, and then the dynamic fracture closure process follows the algorithm that presented in Fig.6.

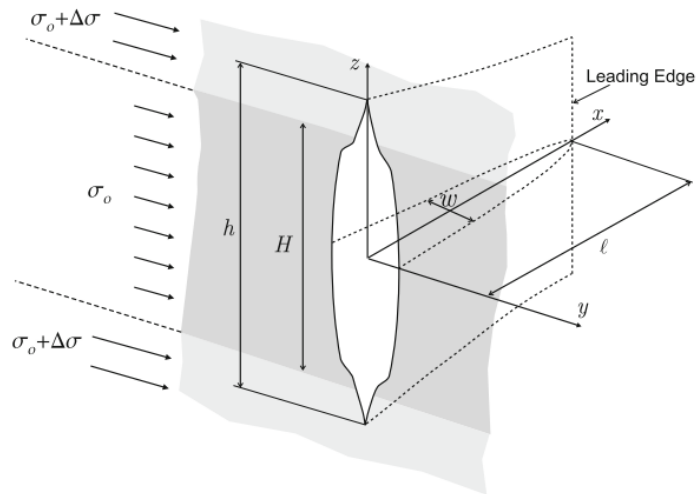

**Fig.18 Illustration of pseudo-3D fracture model (Adachi et al. 2010)**

Despite the existence of abundant literature on P3D fracture propagation, few studies have investigated fracture closure behavior in multi-layered formations. To demonstrate the capability of our model, a case with a fracture penetrating three rock layers with different far-field stresses will be presented. **Fig.19** shows our example case where the fracture is contained in three layers, and both of the top and bottom layer has a higher far-field stress than the middle layer. The total fracture height is 20 m and all the other input parameters remain the same as the previous Base Case.

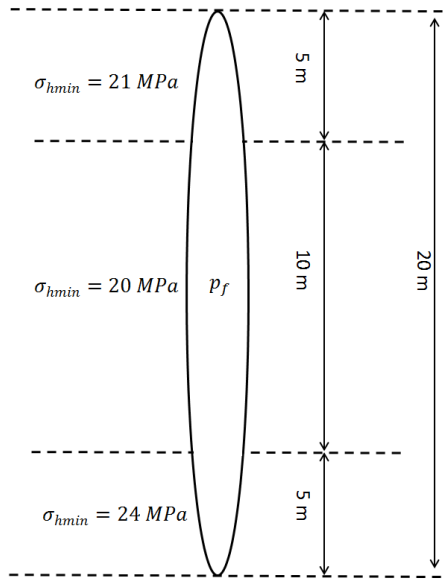

**Fig.19 Illustration of fracture and rock layers with different far-field stress.**

**Fig.20** shows the fracture width distribution across three layers and the corresponding contact stress at different internal fluid pressure. Again, we can notice that at relative high pressure, contact stresses are only concentrated at the fracture edges (top and bottom) and the shape of fracture width is roughly elliptic. As pressure continues to decline, more of the fracture surface comes into contact and the fracture width distribution becomes more influenced by far-field and contact stresses. As expected, the fracture has a larger width in the middle layer where the far-field stress is the smallest, and a smaller fracture width in the bottom layer, where the far-field stress is the highest. Because the contact stress is closely coupled with fracture width, the bottom layer has the highest contact stress. **Fig.21** shows the resulting evolution of fracture average width. There is clear departure from the linear relationship between fracture average width and fluid pressure when the fluid pressure drops close to 24 MPa, which is far-field stress in the bottom layer.

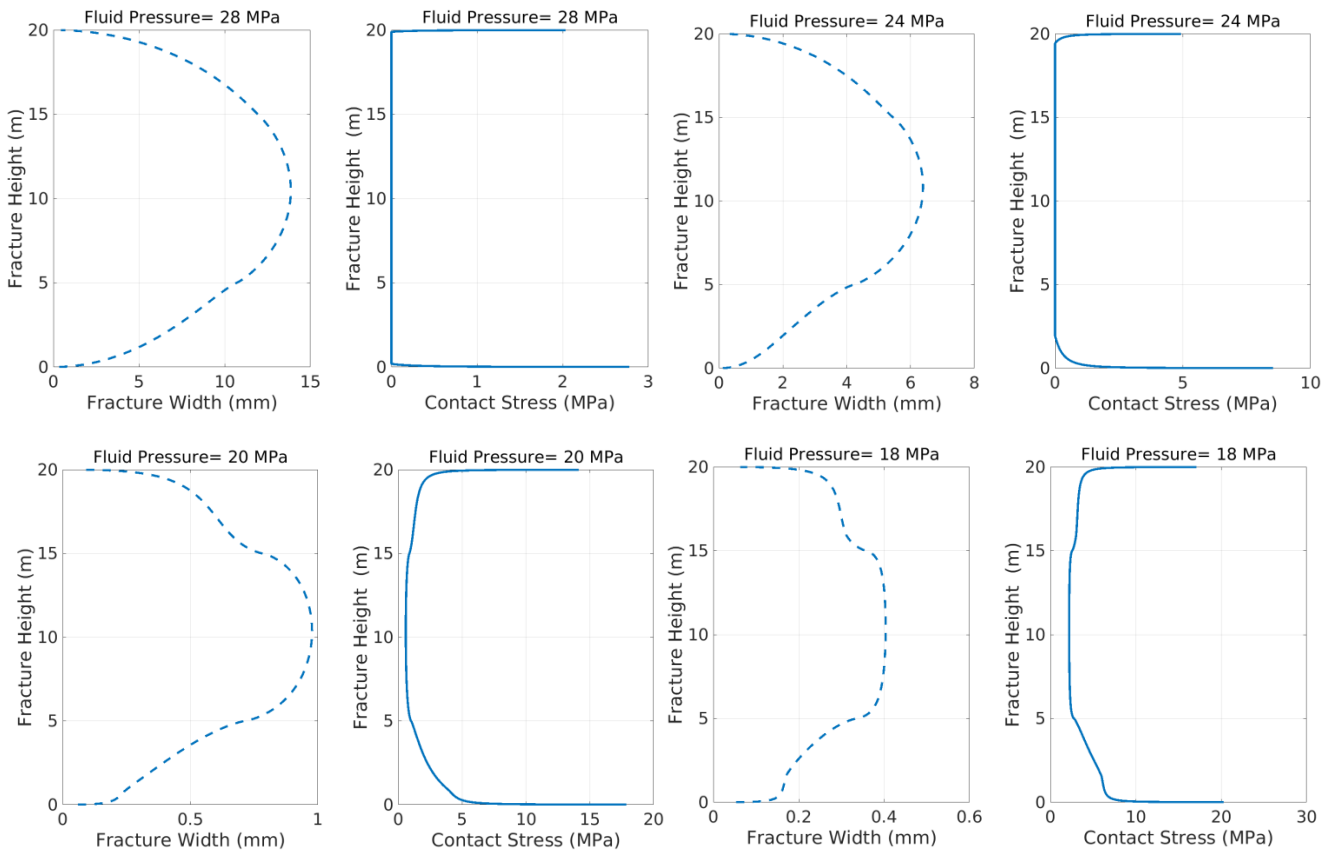

**Fig.20 Fracture width and the corresponding contact stress distribution at different fluid pressure under a non-uniform far-field stress.**

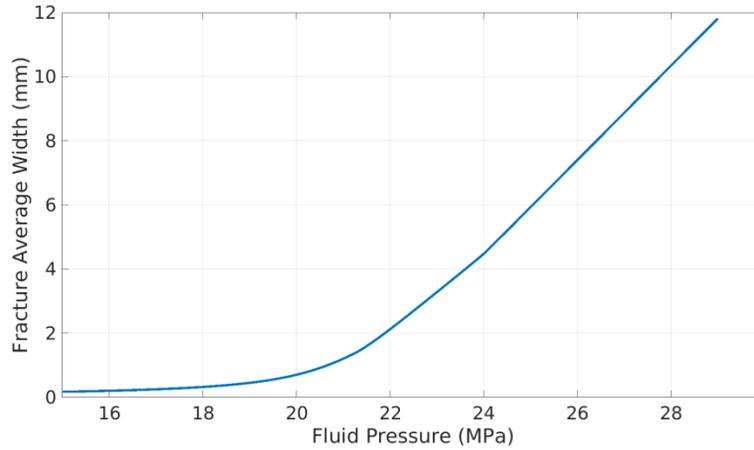

**Fig.21 Fracture average width evolution for different during closure under non-uniform far-field stress.**

## 6. 2  3-D Fracture Closure

Radial or penny-shaped fractures are often created if the fracture is only propagating in a single, relative thick rock layer or the stress difference between the penetrated rock layers is small. The algorithm for modeling dynamic fracture closure behavior presented in Fig.8 is a general approach. To model radial fracture closure, the fracture width distribution under an arbitrary normal load is calculated using Eq.(33). Assuming a radial fracture with a radius of 20 m, and all the other input parameters remaining the same as the previous Base Case, **Fig.22** and **Fig.23** show the evolution of fracture average width for different contact width $w_0$ and contact reference stress $\sigma_{ref}$, respectively. Similar to the results obtained for a 2D fracture geometry, larger contact width leads to earlier departure from the linear relationship between fracture average width and fluid pressure. Also the contact reference stress strongly affects the fracture width evolution after a certain portion of the fracture surface already in contact, but has less influence on the pressure at which the contacting asperities begin to influence the overall fracture closure behavior.

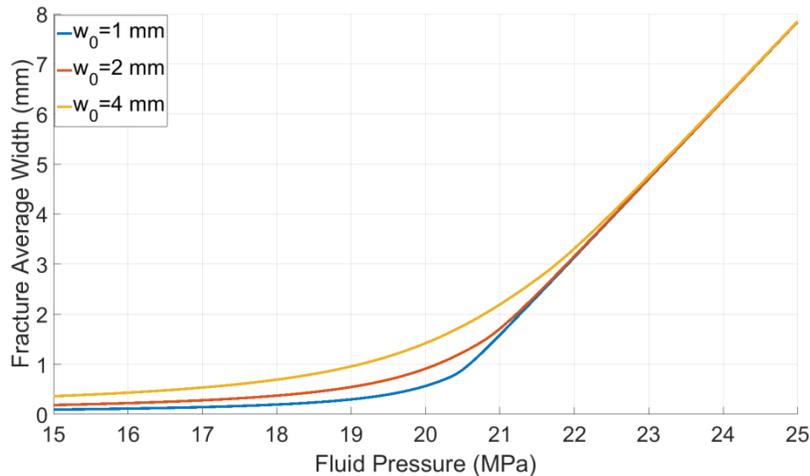

**Fig.22 Fracture average width evolution for different $w_0$ during closure with 3D fracture geometry.**

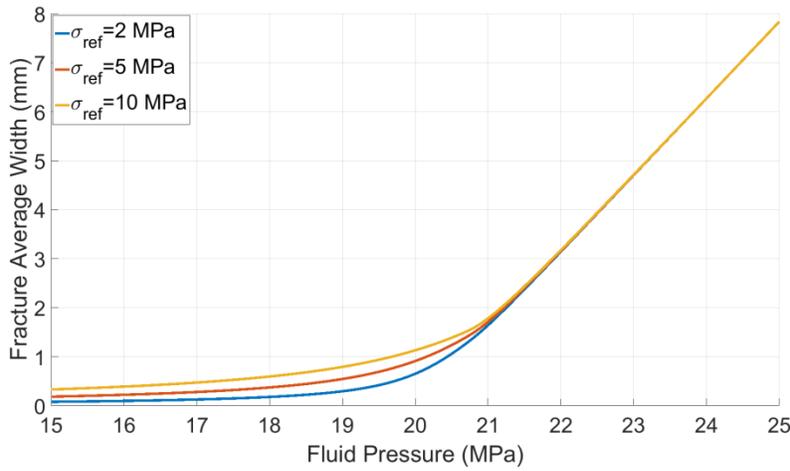

Fig.23 Fracture average width evolution for different $\sigma_{ref}$ during closure with 3D fracture geometry.

## 7. Comparisons of Computation Efficiency

To demonstrate the computational advantage of using the superposition method to determine fracture surface displacement and dynamic fracture closure, the computation speed for both 2D and 3D fracture problems are examined using both the superposition method and the piecewise integral transform (**Appendix**). Unlike the superposition method, the integral transform method calculates fracture surface displacement through nested integration operators. **Fig.24** shows the fracture width profile calculated from these two different methods for 2D and 3D fractures using the parameters from the previous validation cases. The results show that even with a relatively small number of discretized segments (i.e., the fracture length or radius of 50 m is divided into 10 segments), both superposition and integral transform methods provide accurate fracture width profile predictions.

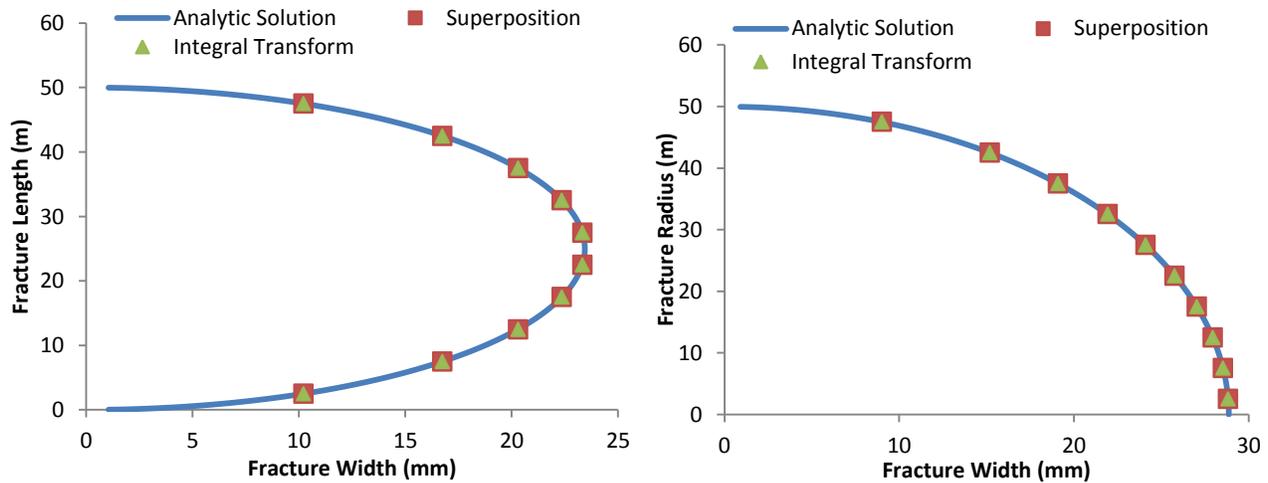

**Fig.24 Comparison of fracture width calculation for different methods with 10 discretized segments**

To test the computational efficiency of superposition and integral transform methods, dynamic fracture closure cases under uniform far-field stress are simulated. For a 2D fracture, the Base Case from Section 6.1.1 is used. For a 3D fracture, the radial case from section 6.2 is used. In all scenarios, the internal pressure drops from 25 MPa to 15 MPa with pressure intervals of 0.1 MPa. **Fig.25** and **Fig.26** show the total computation time needed to complete each simulation case for 2D and 3D fracture closure, respectively, with different numbers of discretized segments. The results indicate that the total computation time and the total number of discretized segments follows a linear relationship on a log-log plot, regardless of which method is used. In addition, we can observe that the superposition method dramatically improved the computational speed for both 2D and 3D cases (by 3 to 4 orders of magnitude).

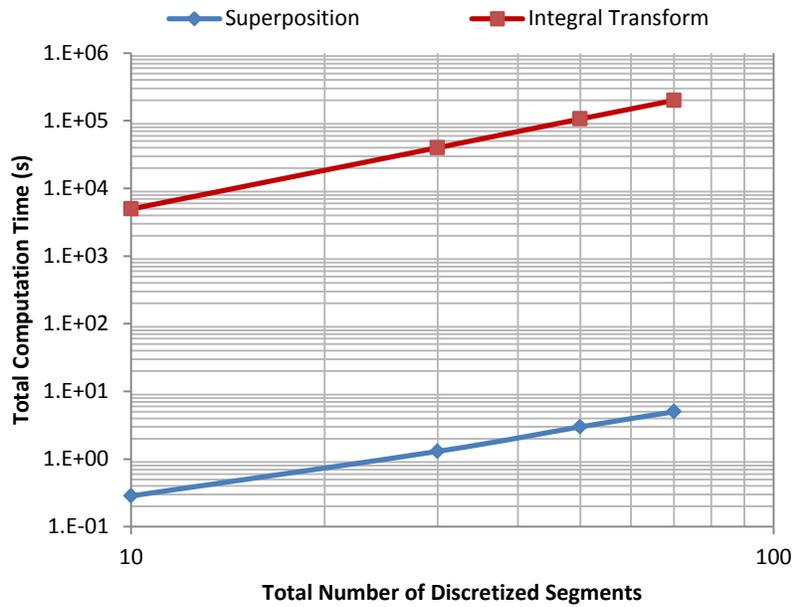

Fig.25 Computation speed comparison for 2D fracture closure process

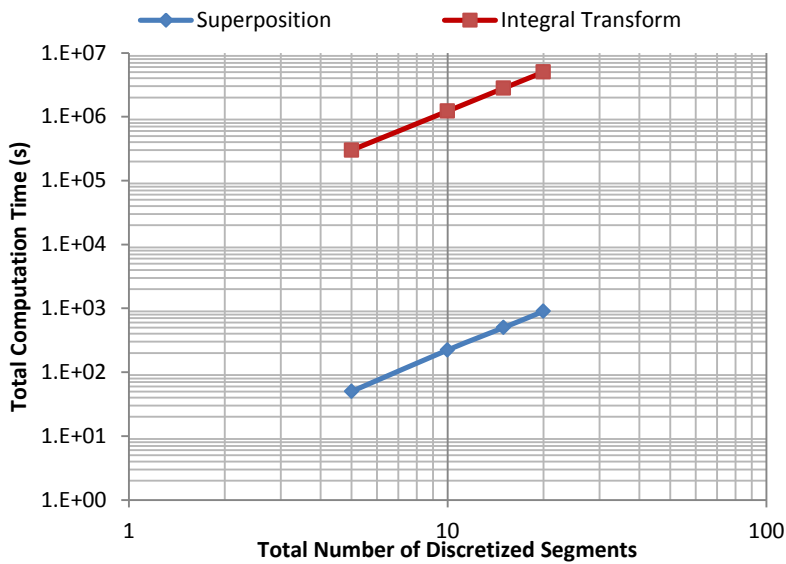

Fig.26 Computation speed comparison for 3D fracture closure process

From our previous analysis, we know that the superposition method yields accurate results even with relatively coarse discretization. However, to better capture dynamic fracture closure behavior and the sharp contact stress profiles, finer discretization may be needed. So to make our dynamic fracture closure model more suitable for handling large scale fractures, parallel computation has been implemented. To test the benefits of parallelization, both 2D and 3D cases are simulated on a workstation with 14 CPUs and the results are presented in **Fig.27**. For the 3D cases, we notice that parallelization reduces computation time for all cases, but the difference becomes smaller when the total number of discretized segments drops below 50. When the total number of discretized segments is above 500, parallelization increases 3D computation speed roughly 10 times. However, for 2D cases, parallelization has a completely different impact on computation efficiency. As can be observed, parallelization increases computation speed only when the total number of discretized segments is above 500, and the more the total number of discretized segments, the more the speedup that can be achieved with parallelization. When the total number of discretized segments is below 500, it actually takes more time to complete a 2D fracture closure simulation with parallelization than just using a single CPU. This is because calculating 2D fracture geometry is relative fast for a small number of discretized segments, and under such circumstances, the amount of time required to coordinate parallel tasks between different CPUs (message passing overhead) surpasses that of doing useful calculations, so parallelization becomes detrimental and increases the total computation time. One way to mitigate this issue is to optimize the number of parallelized CPUs based on the total number of discretized segments through trial and error, but in general, using a single CPU is adequate for a 2D fracture closure problem. Overall, the dynamic fracture closure model (DFCM) presented in this study provides us an approach to model fracture closure process accurately and efficiently. By combining this superposition method with parallel computation, even large scale fracture closure and contact problems can be successfully simulated within minutes.

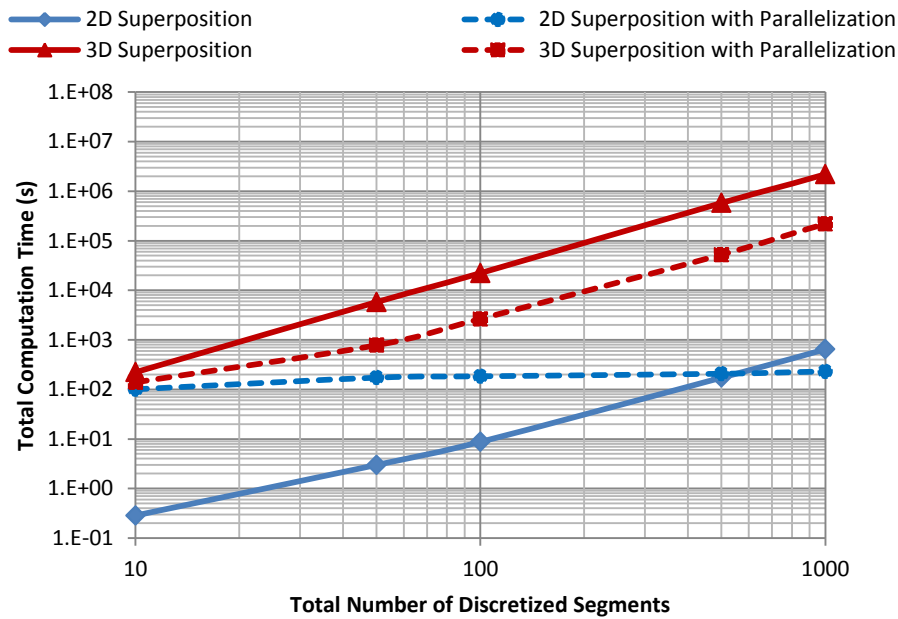

Fig.27 Computation speed comparison with and without parallelization.

## 8. Conclusions

Understanding the role of stress and loading conditions on fracture geometry and the impact of fracture surface asperities and roughness is fundamental to solving fracture closure and contact problems. A wide range of applications, such as underground $CO_2$ or waste repositories, stimulating hydrocarbon reservoirs, geothermal energy exploitation, measurement of minimum in-situ stress from fracture injection test are impacted by such fracture closure problem solutions. The shape of fractures under certain loads in an elastic medium and the corresponding distribution of stress have been well studied in the literature, however, the fracture closure process during unloading has not been investigated thoroughly. Fracture closure problems are challenging because one needs to capture both large scale fracture deformation and small scale contact deformation. In addition, the non-linear nature of contact deformation problems poses serious convergence issues in iterative solutions. These issues make the use of fully-coupled numerical models to simulate fracture closure computationally prohibitive.

In this study, the fracture surface displacements under arbitrary normal load are derived using superposition method. A dynamic fracture closure model (DFCM) for both 2D and 3D fracture geometry is proposed. This model is capable of simulating the fracture closure process due to fluid depressurization under the influence of rough fracture walls and asperities. When compared with previous work using integral transforms, the new model dramatically improves the computational speed (1000 to 10,000 fold increase in speed) with no compromise on accuracy. With the aid of parallel computation, large scale fracture closure and contact problems can now be successfully simulated using our proposed model within a very reasonable time frame.

**Appendix: Fracture Surface Displacement with Arbitrary Normal Load using Nested Integral Transform**

England and Green (1963) showed that the width of a 2D fracture with any arbitrary pressure/stress profile is given by:

$$w_f(y) = \frac{16}{E'} \int_{|y|}^{a} \frac{F(\gamma) + yG(\gamma)}{\sqrt{\gamma^2 - y^2}} d\gamma \tag{A1}$$

where $\gamma$ is a dummy variable, y is the distance to the middle of the fracture, $a$ is the half fracture height (PKN model) or half fracture length (KGD model). Let $P_{Net}(y)$ be the net pressure/stress distribution along the y-direction, which is the total pressure/stress acting on the fracture surfaces minus the far field stress in the fracture opening direction. To obtain $F(\gamma)$ and $G(\gamma)$, $P_{Net}(y)$ has to be divided into even and odd functions, such that $P_{Net}(y) = -f(y) - g(y)$, with f(y) being even and g(y) being odd. Then $F(\gamma)$ and $G(\gamma)$ are determined by:

$$F(\gamma) = -\frac{\gamma}{2\pi} \int_0^\gamma \frac{f(\beta)}{\sqrt{\gamma^2 - \beta^2}} d\beta \tag{A2}$$

$$G(\gamma) = -\frac{1}{2\pi\gamma} \int_0^\gamma \frac{\beta g(\beta)}{\sqrt{\gamma^2 - \beta^2}} d\beta \tag{A3}$$

where $\beta$ is a dummy variable. If the $P_{Net}(y)$ is symmetric to the middle of the fracture, then $g(y)$ and $G(\gamma) = 0$. For example, assuming a uniform pressure $P_{Net}$ inside the fracture, $f(y) = -P_{Net}$, $g(y) = 0$, From Eq.(A2):

$$F(\gamma) = -\frac{\gamma}{2\pi} \int_0^\gamma \frac{-P_{Net}}{\sqrt{\gamma^2 - \beta^2}} d\beta = \frac{\gamma P_{Net}}{4} \tag{A4}$$

Substituting Eq.(A4) into Eq.(A1), we have,

$$w_f(y) = \frac{16}{E'} \int_{|y|}^a \frac{\gamma P_{net}}{4\sqrt{\gamma^2 - y^2}} d\gamma = \frac{4 P_{net}}{E'} \sqrt{a^2 - y^2} \tag{A5}$$

Eq. (A5) is the well-known equation for a 2D static pressurized crack. If a is the fracture half-height, then Eq.(A5) describes the fracture width distribution for the PKN fracture geometry, and if a is the fracture half-length, Eq.(A5) describes the fracture width distribution for the KGD fracture geometry. When an analytic form of $P_{Net}(y)$ is not available a priori (e.g., coupled with other physics) or $P_{Net}(y)$ is not a continuous function, numerical integration to calculate the fracture width has to be implemented in a discretized manner along the fracture height or length. Assume a fracture half-height or half-length is discretized into a number of n segments. In each segment, there exists a uniform net pressure/stress that acts on the fracture surfaces. Then the function of $F(\gamma)$ in the $m^{th}$ segment with net pressure $P_{Netm}$ is calculated as:

$$F(\gamma)_m = -\frac{\gamma}{2\pi} \left( \int_{a_1}^{a_2} \frac{-P_{Net1}}{\sqrt{\gamma^2 - \beta^2}} d\beta + \int_{a_2}^{a_3} \frac{-P_{Net2}}{\sqrt{\gamma^2 - \beta^2}} d\beta + \cdots + \int_{a_m}^{\gamma} \frac{-P_{Netm}}{\sqrt{\gamma^2 - \beta^2}} d\beta \right), \quad a_m \leq y \leq a_{m+1} \tag{A6}$$

Similarly, the integration to obtain $w_f(y)$ also needs to be integrated piecewise. The function of $w_f(y)$ in the $m^{th}$ segment is calculated as:

$$w_f(y)_m = \frac{16}{E'} \left( \int_y^{a_{m+1}} \frac{F(\gamma)_m}{\sqrt{\gamma^2 - y^2}} d\gamma + \int_{a_{m+1}}^{a_{m+2}} \frac{F(\gamma)_{m+1}}{\sqrt{\gamma^2 - y^2}} d\gamma + \cdots + \int_{a_n}^{a_{n+1}} \frac{F(\gamma)_n}{\sqrt{\gamma^2 - y^2}} d\gamma \right), \quad a_m \leq y \leq a_{m+1} \tag{A7}$$

Substitute Eq.(A6) into Eq.(A7):

$$w_f(y)_m = \frac{8}{\pi E'} \int_y^{a_{m+1}} \left( \int_{a_1}^{a_2} \frac{\gamma P_{Net1}}{\sqrt{\gamma^2 - \beta^2}\sqrt{\gamma^2 - y^2}} d\beta + \int_{a_2}^{a_3} \frac{\gamma P_{Net2}}{\sqrt{\gamma^2 - \beta^2}\sqrt{\gamma^2 - y^2}} d\beta + \cdots + \int_{a_m}^{\gamma} \frac{\gamma P_{Netm}}{\sqrt{\gamma^2 - \beta^2}\sqrt{\gamma^2 - y^2}} d\beta \right) d\gamma$$

$$+ \int_{a_{m+1}}^{a_{m+2}} \left( \int_{a_1}^{a_2} \frac{\gamma P_{Net1}}{\sqrt{\gamma^2 - \beta^2}\sqrt{\gamma^2 - y^2}} d\beta + \int_{a_2}^{a_3} \frac{\gamma P_{Net2}}{\sqrt{\gamma^2 - \beta^2}\sqrt{\gamma^2 - y^2}} d\beta + \cdots + \int_{a_{m+1}}^{\gamma} \frac{\gamma P_{Net(m+1)}}{\sqrt{\gamma^2 - \beta^2}\sqrt{\gamma^2 - y^2}} d\beta \right) d\gamma + \cdots$$

$$+ \int_{a_n}^{a_{n+1}} \left( \int_{a_1}^{a_2} \frac{\gamma P_{Net1}}{\sqrt{\gamma^2 - \beta^2}\sqrt{\gamma^2 - y^2}} d\beta + \int_{a_2}^{a_3} \frac{\gamma P_{Net2}}{\sqrt{\gamma^2 - \beta^2}\sqrt{\gamma^2 - y^2}} d\beta + \cdots \right.$$

$$\left. + \int_{a_n}^{\gamma} \frac{\gamma P_{Netn}}{\sqrt{\gamma^2 - \beta^2}\sqrt{\gamma^2 - y^2}} d\beta \right) d\gamma \tag{A8}$$

We define the integration operator I:

$$I(L_\gamma, U_\gamma, L_\beta, U_\beta, P_{Net}) := \frac{8}{\pi E'} \int_{L_\gamma}^{U_\gamma} \int_{L_\beta}^{U_\beta} \frac{\gamma P_{Net}}{\sqrt{\gamma^2 - \beta^2}\sqrt{\gamma^2 - y^2}} d\beta d\gamma \qquad (A9)$$

Re-write Eq.(A8) using the definition from Eq.(A9), we have

$$w_f(y)_m = [I(y, a_{m+1}, a_1, a_2, P_{Net1}) + I(y, a_{m+1}, a_2, a_3, P_{Net2}) + \cdots + I(y, a_{m+1}, a_m, \gamma, P_{Netm})] +$$

$$[I(a_{m+1}, a_{m+2}, a_1, a_2, P_{Net1}) + I(a_{m+1}, a_{m+2}, a_2, a_3, P_{Net2}) + \cdots + I(a_{m+1}, a_{m+2}, a_{m+1}, \gamma, P_{Net(m+1)})] +$$

$$\cdots + [I(a_n, a_{n+1}, a_1, a_2, P_{Net1}) + I(a_n, a_{n+1}, a_2, a_3, P_{Net2}) + \cdots + I(a_n, a_{n+1}, a_n, \gamma, P_{Netn})] \qquad (A10)$$

Notice that $w_f(y)_m$ is the summation of the integration operator I that operated on different bounds and integrands, we put all the terms in the R.H.S of Eq.(A10) in to a matrix, called MatrixI(k,j), with k rows and j columns. Then Eq.(A10) becomes

$$w_f(y)_m = \sum \text{MatrixI}(k, j) \qquad (A11)$$

Re-arrange MatrixI, we have

$$w_f(y)_m$$
$$= \sum \begin{bmatrix} I(y, a_{m+1}, a_1, a_2, P_{Net1}) & I(y, a_{m+1}, a_2, a_3, P_{Net2}) & \cdots & I(y, a_{m+1}, a_m, \gamma, P_{Netm}) \\ I(a_{m+1}, a_{m+2}, a_1, a_2, P_{Net1}) & I(a_{m+1}, a_{m+2}, a_2, a_3, P_{Net2}) & \cdots & I(a_{m+1}, a_{m+2}, a_{m+1}, \gamma, P_{Net(m+1)}) \\ \vdots & \vdots & \vdots & \vdots \\ I(a_n, a_{n+1}, a_1, a_2, P_{Net1}) & I(a_n, a_{n+1}, a_2, a_3, P_{Net2}) & \cdots & I(a_n, a_{n+1}, a_n, \gamma, P_{Netn}) \end{bmatrix} \qquad (A12)$$

Based on Eq.(A12), an algorithm can be developed to calculated fracture width within the $m^{th}$ segment of a 2D fracture, and this process has to be repeated from the $1^{st}$ segment to the $n^{th}$ segment to obtain the entire fracture width profile.

The radial crack opening due to an excess pressure distribution in a linearly elastic material of infinite extent can be expressed as (Sneddon and Lowengrub 1969):

$$w_f(r_D) = \frac{8R_f}{\pi E'} \int_{r_D}^{1} \frac{du}{\sqrt{u^2 - r_D^2}} \int_0^u \frac{sP_{net}(s)ds}{\sqrt{u^2 - s^2}} \qquad (A13)$$

where $R_f$ is the length of fracture radius, u and s are dummy variables and $r_D$ is the normalized radius, which is defined as

$$r_D = \frac{r}{R_f} \qquad (A14)$$

where r is the local fracture radius. Assume a fracture is discretized into a number of n segments along radial direction, if $r$ belongs to the $m^{th}$ segment where $a_m \leq r \leq a_{m+1}$, then

$$w_f(r_D)_m = \frac{8R_f}{\pi E'} \left\{ \left[ \int_{\frac{a_1}{R}}^{\frac{a_2}{R}} \frac{sP_{Net1}}{r_D} F\left(\sin^{-1}\sqrt{\frac{1-r_D^2}{1-s^2}}, \frac{s}{r_D}\right) ds + \int_{\frac{a_2}{R}}^{\frac{a_3}{R}} \frac{sP_{Net2}}{r_D} F\left(\sin^{-1}\sqrt{\frac{1-r_D^2}{1-s^2}}, \frac{s}{r_D}\right) ds + \cdots + \right.\right.$$

$$\left. \int_{\frac{a_m}{R}}^{r_D} \frac{sP_{Netm}}{r_D} F\left(\sin^{-1}\sqrt{\frac{1-r_D^2}{1-s^2}}, \frac{s}{r_D}\right) ds \right] + \left[ \int_{r_D}^{\frac{a_{m+1}}{R}} P_{Netm} F\left(\sin^{-1}\sqrt{\frac{1-s^2}{1-r_D^2}}, \frac{r_D}{s}\right) ds + \right.$$

$$\left. \int_{\frac{a_{m+1}}{R}}^{\frac{a_{m+2}}{R}} P_{Net(m+1)} F\left(\sin^{-1}\sqrt{\frac{1-s^2}{1-r_D^2}}, \frac{r_D}{s}\right) ds + \cdots + \int_{\frac{a_n}{R}}^{\frac{a_{n+1}}{R}} P_{Netn} F\left(\sin^{-1}\sqrt{\frac{1-s^2}{1-r_D^2}}, \frac{r_D}{s}\right) ds \right]\right\} \quad (A15)$$

We define integration operators R1 and R2:

$$R1(L_s, U_s, P_{Net}) := \frac{8R_f}{\pi E'} \int_{L_s}^{U_s} \frac{s P_{Net}}{r_D} F\left(\sin^{-1}\sqrt{\frac{1-r_D^2}{1-s^2}}, \frac{s}{r_D}\right) ds \quad (A16)$$

$$R2(L_s, U_s, P_{Net}) := \frac{8R_f}{\pi E'} \int_{L_s}^{U_s} P_{Net} F\left(\sin^{-1}\sqrt{\frac{1-s^2}{1-r_D^2}}, \frac{r_D}{s}\right) ds \quad (A17)$$

We can rewrite Eq.(A15) using Eq.(A16) and (A17), to get:

$$w_f(r_D) = \left[ R1\left(\frac{a_1}{R_f}, \frac{a_2}{R_f}, P_{Net1}\right) + R1\left(\frac{a_2}{R_f}, \frac{a_3}{R_f}, P_{Net2}\right) + \cdots + R1\left(\frac{a_2}{R_f}, r_D, P_{Netm}\right) \right]$$

$$+ \left[ R2\left(r_D, \frac{a_{m+1}}{R_f}, P_{Netm}\right) + R2\left(\frac{a_{m+1}}{R_f}, \frac{a_{m+2}}{R_f}, P_{Net(m+1)}\right) + \cdots R2\left(\frac{a_n}{R_f}, \frac{a_{n+1}}{R_f}, P_{Netn}\right) \right] \quad (A18)$$

From Eq.(A18), we can observe that $w_f(r_D)$ can be expressed by the summation of two 1D matrices:

$$w_f(r_D)_m = \sum MatrixR1 + \sum MatrixR2 \quad (A19)$$

where

$$MatrixR1 = \left[ R1\left(\frac{a_1}{R_f}, \frac{a_2}{R_f}, P_{Net1}\right) \; R1\left(\frac{a_2}{R_f}, \frac{a_3}{R_f}, P_{Net2}\right) \; \cdots \; R1\left(\frac{a_2}{R_f}, r_D, P_{Netm}\right) \right] \quad (A20)$$

$$MatrixR2 = \left[ R2\left(r_D, \frac{a_{m+1}}{R_f}, P_{Netm}\right) \; R2\left(\frac{a_{m+1}}{R_f}, \frac{a_{m+2}}{R_f}, P_{Net(m+1)}\right) \; \cdots \; R2\left(\frac{a_n}{R_f}, \frac{a_{n+1}}{R_f}, P_{Netn}\right) \right] \quad (A21)$$

Based on Eq.(A19) to Eq.(A21), an algorithm can be developed to calculated fracture width within the $m^{th}$ segment of a radial fracture, and this process has to be repeated from the $1^{st}$ segment to the $n^{th}$ segment to obtain the entire fracture width profile.